\documentclass[a4paper,11pt]{article}
\pdfoutput=1 

\usepackage{jinstpub} 
\usepackage{array}
\usepackage{gensymb}

\title{Full Scale Proton Beam Impact Testing of new CERN Collimators and Validation of a Numerical Approach for Future Operation}

\author[1]{M. Bergeret,\note{Corresponding author.}}
\author[]{F-X. Nuiry,}
\author[]{M. Calviani,}
\author[]{M.A. Fraser,}
\author[]{M. Butcher,}
\author[]{L-M. Grec,}
\author[]{L. Gentini,}
\author[]{A. Lechner,}
\author[]{M. Frankl,}
\author[]{V. Rizzoglio,}
\author[]{S. Burger,}
\author[]{A. Cherif}

\affiliation[]{CERN,\\1211 Geneva 23, Switzerland}

\emailAdd{maxime.pierre.jean.bergeret@cern.ch}

\abstract{
New collimators are being produced at CERN in the framework of a large particle accelerator upgrade project to protect beam lines against stray particles. Their movable jaws hold low density absorbers with tight geometric requirements, while being able to withstand direct proton beam impacts.
Such events induce considerable thermo-mechanical loads, leading to complex structural responses, which make the numerical analysis challenging. Hence, an experiment has been developed to validate the jaw design under representative conditions and to acquire online results to enhance the numerical models. 
Two jaws have been impacted by high-intensity proton beams in a dedicated facility at CERN and have recreated the worst possible scenario in future operation. The analysis of online results coupled to post-irradiation examinations have demonstrated that the jaw response remains in the elastic domain. However, they have also highlighted how sensitive the jaw geometry is to its mounting support inside the collimator. Proton beam impacts, as well as handling activities, may alter the jaw flatness tolerance value by $\pm$ 70 \textmu m, whereas the flatness tolerance requirement is 200 \textmu m. In spite of having validated the jaw design for this application, the study points out numerical limitations caused by the difficulties in describing complex geometries and boundary conditions with such unprecedented requirements.}


\arxivnumber{XXX.YYYY} 

\begin{document}
\maketitle
\flushbottom

\section{Introduction}
\label{sec:sec1}
CERN is the European Organization for Nuclear Research located in Geneva on the border between France and Switzerland. Many activities at CERN currently involve upgrading the world's largest particle accelerator complex to increase the luminosity of the Large Hadron Collider (hereinafter ``LHC''). In that framework, a large improvement of the LHC injector chain is being conducted under the LHC Injector Upgrade Project (hereinafter ``LIU Project'')~\cite{1}. The proton beams for the LHC are fast-extracted from the last injector, the Super Proton Synchrotron (hereinafter ``SPS''), at 450 GeV/c through two transfer lines: TI 2 and TI 8. A machine protection system had been developed at the time of the LHC's construction to protect the beam line equipment against possible proton beam steering errors. By means of collimators, it was intended to attenuate the damage potential of the proton beams, which is related to the intensity and emittance, below a safe level corresponding to the damage limit of accelerator components. The LIU requirements have involved the increase of the attenuation capacity of the so-called TCDI collimators (``Target Collimator Dump Injection'') by a factor of 3.5, and has consequently required the construction of a new hardware: the TCDIL collimators (``Target Collimator Dump Injection Long'').

\subsection{TCDIL collimators}
\label{sec:sec1pt1}
The TCDI collimators employed isostatic graphite absorber blocks (density = 1.8 g.cm$^{-3}$) over a length of 1.2 m to intercept off-trajectory proton beams. The TCDIL collimators require the installation of low density absorber blocks (density > 1.7 g.cm$^{-3}$) over a length of 2.1 m, for which material selection has been driven by a beam impact testing experiment~\cite{2}. 

\begin{figure}[htbp]
\centering 
\includegraphics[width=0.9\textwidth,trim=0 0 0 0,clip]{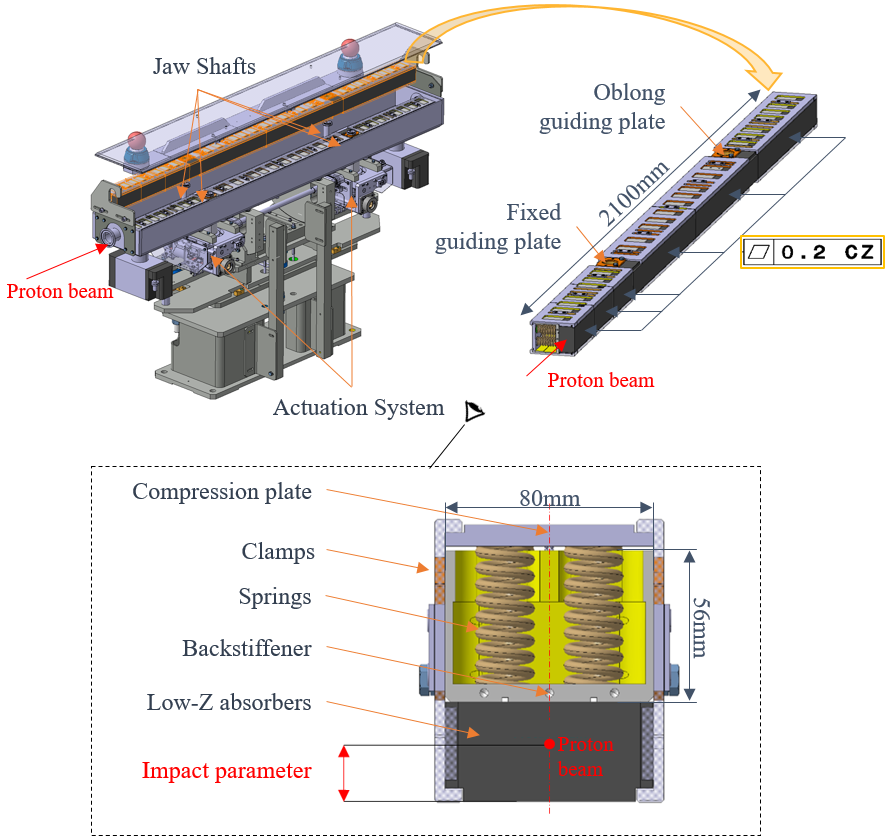}
\qquad
\caption{\label{fig:1} Exploded view of a horizontal TCDIL collimator showing the jaws linked to the actuation system by means of cylindrical shafts. Detailed view of the jaw by outlining the assembly parts and the 200 \textmu m flatness requirement over the common active absorber surfaces.}
\end{figure}

As shown in Figure~\ref{fig:1}, each collimator contains a pair of sub-mechanical assemblies (hereinafter ``jaws''), which hold the 2.1 m-long series of absorbing blocks. A flatness tolerance of 200 \textmu m is requested over the common inner absorber plane for precise positioning around the proton beams~\cite{3}. To do so, the jaws employ two patterns: A 2.1 m-long grade 316L stainless steel stiffener (hereinafter ``backstiffener'') that governs the overall jaw geometry, and a grade 304L stainless steel clamping system with compression springs, clamps and compression plates, that holds the absorbing blocks on the backstiffener. The jaws are then installed in an ultra-high vacuum tank on two cylindrical shafts going through a set of fixed and oblong GLIDCOP\textregistered\ AL-15 guiding plates, which are linked to an actuation system with four independent translational degrees-of-freedom. The jaws can therefore translate in the transverse direction with adjustment of their angular orientation with respect to the beam trajectory. A static vacuum pressure of $5.0\times 10^{-9}$ mbar is required with a maximum total ougassing flux of $2.0\times 10^{-7}$ mbar.L.s$^{-1}$. Six collimators will be installed at specific positions in each of the two transfer lines TI 2 and TI 8.

\subsection{Jaw design requirements}
\label{sec:sec1pt2}
The jaw design is challenging as it must withstand direct impacts of LIU proton beams without jeopardizing the 200 \textmu m flatness tolerance requirement. Table~\ref{tab:1} states the characteristics of the standard LIU proton beam. An extensive analysis~\cite{3} studied the different accidental scenarios causing beam mis-steering, and has concluded that the jaw absorbers would intercept proton beams between 0 mm and 6.9 mm from the inner absorber plane. This distance is named the impact parameter, as seen in Figure~\ref{fig:1}. 

In case of a direct proton beam impact, the particles interact with the absorber nuclei to locally generate extreme heat. In addition, the primary nuclear interactions spread a shower of secondary particles that deposits additional energy in the different stainless steel parts of the assembly. Such a fleeting event, in the order of microseconds, induces a thermal burst that turns into a complex structural response with the excitation of many dynamic modes. Numerical models have been defined in ANSYS\textregistered\ to study the jaw structural response, especially in the worst case scenario, and its design studied accordingly. However, despite the increasing accuracy of numerical approaches, uncertainties related to the model assumptions remained. An experimental validation was therefore required.

\begin{table}[htbp]
\centering
\caption{\label{tab:1} Proton beam requirements for standard LIU and maximum proton beam capabilities at the HiRadMat facility. The intensity of a standard LIU beam is 76 \%  higher.}
\smallskip
\begin{tabular}{| l | c| c | }
\hline
\textbf{Beam Parameters} & \textbf{Standard LIU} & \textbf{HiRadMat} \\ 
\hline
Proton energy [GeV] & 450 & 440 \\ 
\hline
Number of bunch & 288 & 288 \\
\hline
Bunch intensity  & $2.3\times 10^{11}$ & $1.3\times 10^{11}$ \\
\hline
Pulse length [\textmu s]  & 7.8 & 7.2 \\
\hline
Horizontal beam size at 1$\sigma$ [mm] & 0.405 & 0.313 \\
\hline
Vertical beam size at 1$\sigma$ [mm] & 0.647 & 0.313 \\
\hline
\end{tabular}
\end{table}

\subsection{Proton beam impact testing}
\label{sec:sec1pt3}
CERN's HiRadMat facility~\cite{4} uses a dedicated experimental area, which can be irradiated by SPS proton beams at a momentum of of 440 GeV/c. The facility is especially convenient thanks to a wide flexibility of beam intensities and beam optic parameters. It is commonly used for research and development of beam intercepting materials and validation of equipment~\cite{2, 41, 42, 43}. Table~\ref{tab:1} shows the maximum HiRadMat irradiation capabilities. 

A HiRadMat beam impact testing experiment was hence developed to assess the jaw design subjected to proton beam impacts as required by the LIU project. The so-called HiRadMat 44 experiment (hereinafter ``HRMT 44''), had to both (i) confirm that the jaw geometry is not severely affected in the worst case scenario, and (ii) provide experimental online results to enhance and validate the numerical models. The experiment is of considerable importance to offer reliable collimator equipment and guide future maintenance strategies. Finally, HRMT 44 was conceived also as an opportunity to (iii) validate the integrity of a new absorbing material as performed in~\cite{2}. The Novoltex\textregistered\ Sepcarb\textregistered\ 054-62 is a 3D carbon-carbon composite manufactured by ArianeGroup and has thermo-mechanical properties similar to the already validated Naxeco\textregistered\ Sepcarb\textregistered\ 358-02~\cite{5}.

This study will present the numerical approach that has been used to analyze the structural response of impacted jaws. The model assumptions will especially be scrutinised. Then, the experimental set-up will be developed with a focus on the instrumentation and its role. Afterwards, the experimental procedure and the online results will be shown for comparison with the numerical results. Finally, the post-irradiation examinations will assess the jaw integrity after being subjected to the proton beam impacts.

\section{Numerical analysis of proton beam impacts in the collimator jaw}
\label{sec:sec2}
The thermo-mechanical description of the interaction between a proton beam and a collimator jaw relies on several physical problems and engineering methodologies. For that purpose, finite element methods, and related software platforms such as ANSYS\textregistered\ Workbench, offer an increasing opportunity to define and solve 3D models with complex geometries, loads and responses. This section examines the jaw structural responses, especially in the worst case scenario, after presenting the required simulation workflow.

\subsection{Simulation workflow}
\label{sec:sec2pt1}
\begin{figure}[htbp]
\centering 
\includegraphics[width=0.9\textwidth,trim=0 0 0 0,clip]{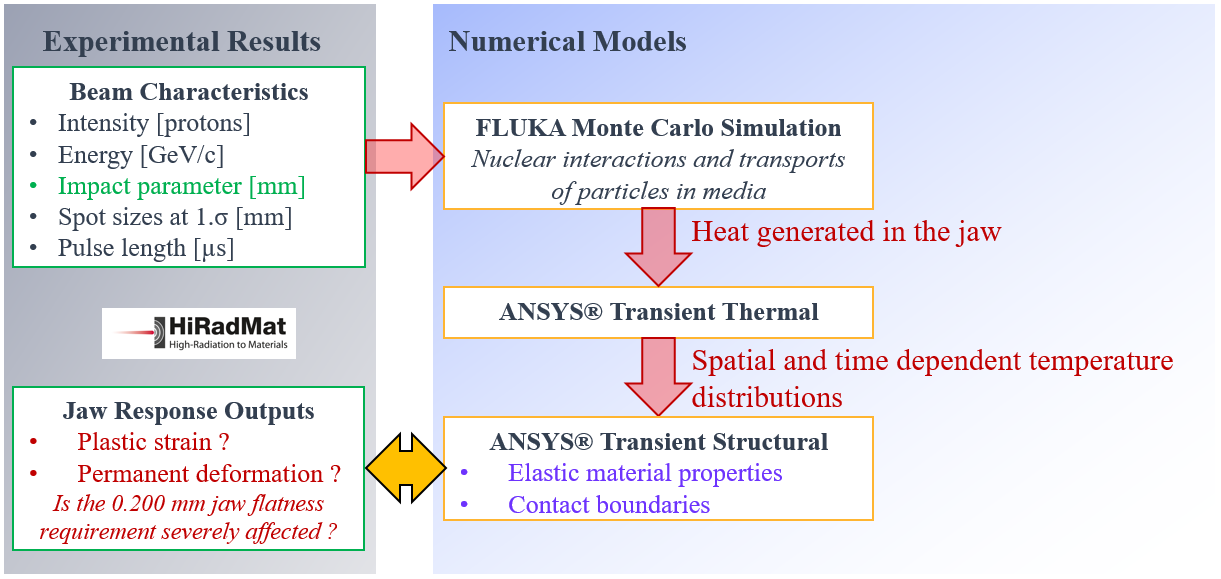}
\qquad
\caption{\label{fig:2} Workflow diagram of the numerical models defined in FLUKA and ANSYS\textregistered\ for studying the jaw thermo-mechanical response in case of a proton beam impact. The principal variable input is the beam impact parameters (in green). The ANSYS\textregistered\ Transient Structural model is based on two main assumptions (in purple) that must be confirmed by the experimental results (in red).}
\end{figure}

As shown in Figure~\ref{fig:2}, the simulation workflow that leads to the structural problem integrates two successive and uncoupled steps. The physical problem, or simulating the nuclear interactions and transports of particles in media, is a prerequisite solved by the FLUKA~\cite{i} Monte Carlo code. It provides the heat density generated in the collimator jaw according to pre-defined beam characteristics. The thermal problem can be solved thereafter by an ANSYS\textregistered\ transient thermal model, which calculates the temperature distribution at the end of the pulse and its evolution related to conduction, convection and radiation processes. The spatial and time dependent temperature distributions can finally be imported in the ANSYS\textregistered\ transient structural model.

The severity of the beam impacts depends on the beam characteristics summarized in Figure~\ref{fig:2}. The intensity, energy and pulse length are defined by the LIU beam requirements. It has been demonstrated that the jaw structural response is not sensitive (within a factor of two) to the beam spot sizes, defined as the standard deviations ($\sigma$) of the beam protons. Smaller beam sizes only induce more focused loads in the absorbers due to the primary particle interactions. Hence, the principal variable input of the simulation workflow is the beam impact parameter. ``Deep impacts'' lead to higher total energy deposition in the stainless steel backstiffener, and consequently, the deepest beam impact parameter (6.9 mm, see section~\ref{sec:sec1pt2}) embodies the highest risk of affecting the jaw integrity.

\subsection{Structural response}
\label{sec:sec2pt2}
\begin{figure}[htbp]
\centering 
\includegraphics[width=1\textwidth,trim=0 0 0 0,clip]{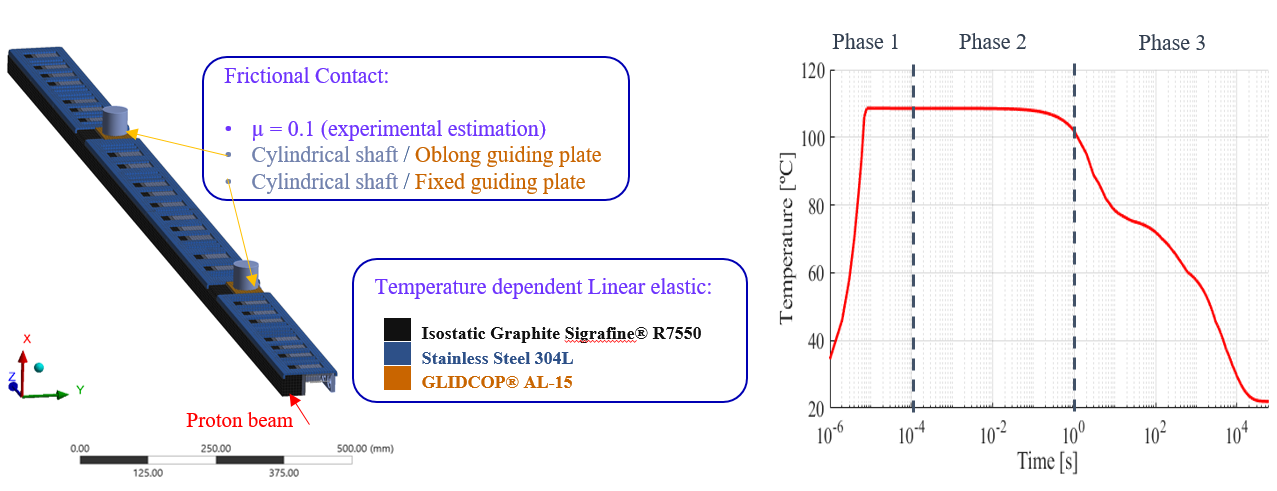}
\qquad
\caption{\label{fig:3} Full scale 3D geometry of the jaw defined in the ANSYS\textregistered\ transient structural model. Two frictional contacts model the connection between the jaw guiding plates and the cylindrical shafts with a friction coefficient of \textmu\ = 0.1. Temperature dependent linear elastic material properties are implemented. Finally, the simulation set up can describe three phases with distinct structural modes, delimited by dotted lines in a characteristic peak temperature evolution in the backstiffener after the ``deepest standard LIU proton beam impact''.}
\end{figure}

The duration of a beam induced heat load is of the order of a few microseconds as it corresponds to the pulse length, as seen in Table~\ref{tab:1}. As shown in Figure~\ref{fig:3}, the backstiffener's peak temperature shortly ramps in 7.8 \textmu s to a quasi-adiabatic thermal state, just like the absorbers and the other stainless steel parts. Such a fleeting event partly prevents thermal expansion because of an inertia effect, and gives rise to a dynamic response that can be thought of as various mode shapes of the jaw being excited with different characteristic times. The model is consequently set up across three phases, delimited by dotted lines in Figure~\ref{fig:3}.

\begin{figure}[htbp]
\centering 
\includegraphics[width=1\textwidth,trim=0 0 0 0,clip]{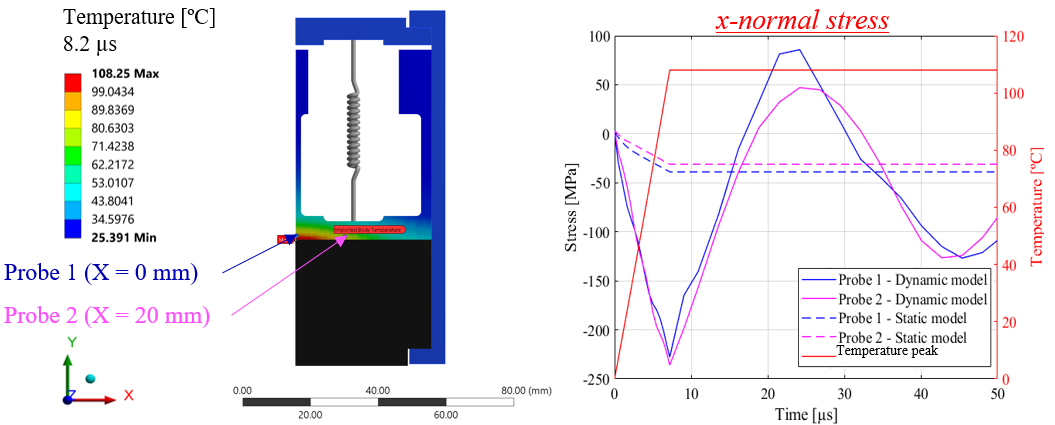}
\qquad
\caption{\label{fig:4} Preliminary 2D-plane-strain model in ANSYS\textregistered\ transient structural of a jaw cross section with the higher backstiffener temperature profile. In case of the deepest standard LIU proton beam impact, the peak temperature in the backstiffener is of 108 \degree C. The right plot shows how the short duration thermal load creates larges compressive-to-tensile x-normal stresses.}
\end{figure}

In the first phase, the purely thermal stresses, or quasi-static stresses created by a non-uniform temperature field, are superposed with high-frequency dynamic modes (with characteristic times of a few microseconds) with elastic and potentially elastic-plastic total stresses in the backstiffener. A preliminary 2D-plane-strain analysis, presented in Figure~\ref{fig:4} in the case of the deepest standard LIU beam impact, had been defined at a characteristic jaw cross-section with the higher temperature profile in the backstiffener. The plane strain assumption was hitherto justified by the jaw longitudinal dimension and by the duration of the heat load that tends to prevent longitudinal expansion~\cite{6}. The plot in Figure~\ref{fig:4} highlights how dynamic the structural response is by comparing the static and transient x-normal stresses at two points. Strong inertia effects are observed with large compressive-to-tensile dynamic stresses, reaching -236 MPa at the end of the beam pulse. From that analysis, the simulation set up uses short time steps able to estimate the high-frequency dynamic modes up to 100 \textmu s. After that time, the dynamic stresses are largely damped and no longer show a risk of resulting in plastic strains.

\begin{figure}[htbp]
\centering 
\includegraphics[width=1\textwidth,trim=0 0 0 0,clip]{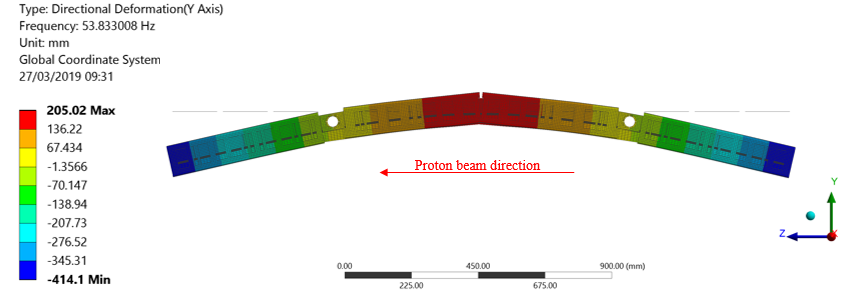}
\qquad
\caption{\label{fig:5} Preliminary ANSYS\textregistered\ modal analysis of the jaw in the thermal state reached at the end of the proton beam pulse. The fundamental has a frequency of 54 Hz and governs the response in the second phase.}
\end{figure}

At a longer time scale, the response is governed by lower frequency modes (with characteristic times of a few milliseconds) with mode shapes that impel large bending oscillations. The structural model has to estimate the maximum deformation amplitudes and ensure that the risk of contact with the ultra-high vacuum tank is limited. An ANSYS\textregistered\ modal analysis, presented in Figure~\ref{fig:5}, had evaluated the fundamental at a frequency of 54 Hz to set up accordingly the simulation time step in the second phase, which lasts for 1 s. The bending oscillations are then largely attenuated as the friction at the contact between the guiding plates and the cylindrical shafts dissipates energy.

\begin{figure}[htbp]
\centering 
\includegraphics[width=1\textwidth,trim=0 0 0 0,clip]{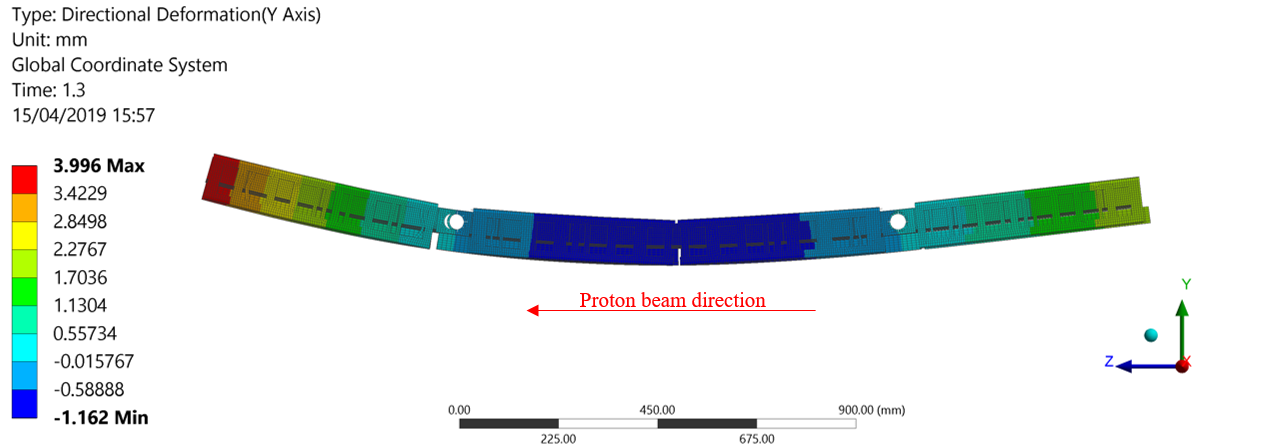}
\qquad
\caption{\label{fig:6} ANSYS\textregistered\ static structural analysis of the jaw response reached after at the beginning of the third  phase (1 s) after the deepest standard LIU proton beam impact. A deformed geometry results from the thermal strains with the jaw downstream and upstream sides extending away from the beam trajectory.}
\end{figure}

Afterwards, the jaw structural response reaches a quasi-static state deformed by thermal strains. As most of the beam energy deposition occurs in the absorber blocks and in the lower part of the backstiffener, the jaw acquires a convex geometry with the downstream and upstream sides extending away from the beam trajectory, as shown in Figure~\ref{fig:6}. Thermal conduction and energy dissipation processes finally govern the last phase until the system has cooled to the initial temperature condition.

\subsection{Structural model assumptions}
\label{sec:sec2pt3}
A full scale 3D jaw geometry has been defined in the ANSYS\textregistered\ transient structural model with a symmetry about the y-z plane. The model is based on two strong assumptions, which are highlighted in purple in Figures~\ref{fig:2} and ~\ref{fig:3} . First, the implemented material properties are temperature dependent linear elastic. That is questionable given the large dynamic stresses discussed in section~\ref{sec:sec2pt2}. Then, two frictional contacts model the connection between the jaw guiding plates and the cylindrical shafts with a friction coefficient of \textmu\ = 0.1. The analysis of the HRMT 44 online results has converged to this value, see section~\ref{sec:sec5pt2}, although it had been previously estimated based on empirical studies. A frictional mount brings on the risk of residual deformations after cooling.

The challenging jaw design requirements, see section~\ref{sec:sec1pt2}, cannot be guaranteed with certainty from a numerical approach, especially considering the jaw dimensions and the design complexity. A HiRadMat beam impact testing experiment was the opportunity to test a jaw under  thermo-mechanical loads representative of LIU accidental scenarios, and ideally, provide an experimental validation. On the other hand, experimental online results could be compared to the numerical results to enhance the accuracy of the model.

\section{HiRadMat 44 set-up}
\label{sec:sec3}

\subsection{Test bench}
\label{sec:sec3pt1}
\begin{figure}[htbp]
\centering 
\includegraphics[width=0.75\textwidth,trim=0 0 0 0,clip]{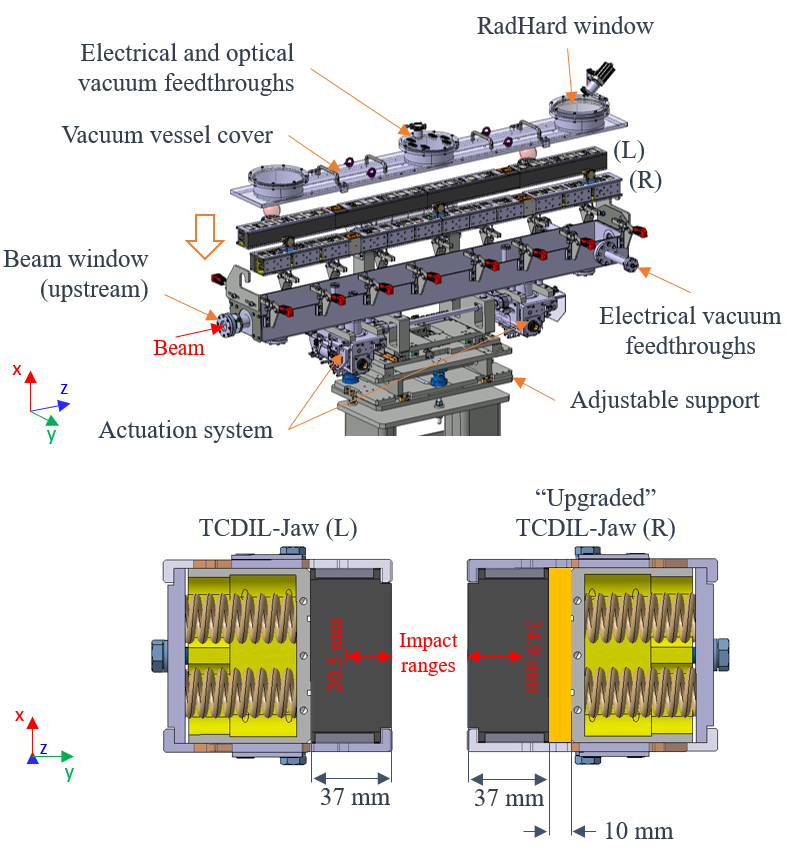}
\qquad
\caption{\label{fig:7} Two jaw designs are mounted inside the HRMT 44 test bench, whose design is largely based on the TCDIL collimator. On the left with respect to the beam direction, a TCDIL jaw with an absorber thickness of 37 mm. On the right, an upgraded TCDIL jaw with a total absorber thickness of 47 mm.}
\end{figure}

The HRMT 44 test bench is based on a prototype TCDIL collimator. As shown in Figure~\ref{fig:7}, two jaw designs are linked to a standard collimator actuation system with a precision and repeatability of $\pm$ 20 \textmu m~\cite{7} controlled by four stepper motors and four Linear Variable Differential Transformer (hereinafter ``LVDT'') position sensors. The jaws are installed in a primary vacuum tank. The cover is closed with fast-clamps to speed up the dismounting operations in view of the post-irradiation examination, hence limiting the personal exposure to ionizing radiation resulting from proton beam impacts. Two 85 mm-diameter proton beam windows made of 0.254 mm-thick PF-60 beryllium foil~\cite{8} ensure the vacuum tightness while inducing a minimum interaction with the beam. Flanges integrate several electrical feedthroughs and a radiation resistant glass window offers a visual port to the jaws. The vacuum vessel is mounted on a finely 3D-adjustable support with a fifth translational degree of freedom controlled by one stepper motor and one LVDT to increase the stroke range of the jaws, mentioned in section~\ref{sec:sec3pt2}. The equipment is finally installed on a standard HiRadMat table, which provides an electrical plug-in interface for a quick installation in the dedicated experimental area. 

\subsection{Collimator jaws}
\label{sec:sec3pt2}
The first jaw design, mounted on the left-hand side (L) in the direction of the beam, is a standard TCDIL jaw. Four 37 mm-thick Sigrafine\textregistered\ R7550 graphite blocks formed the 2.1 m absorber length. The block lengths are either 510 mm or 540 mm depending on their position along the jaw. The second jaw design, mounted on the right-hand side (R), is a slightly modified TCDIL jaw with a total absorber layer of 47 mm instead of 37 mm as shown in Figure~\ref{fig:7}. The additional layer is composed of two 1050 mm-long Sigrafine\textregistered\ R7550 graphite plates. It increases the relative distance between the proton beam and the stainless steel backstiffener, and consequently decreases the beam-induced energy deposition in the backstiffener. The upstream graphite block was replaced by three 170 mm-long Novoltex\textregistered\ Sepcarb\textregistered\ 054-62 blocks to study objective (iii), see section~\ref{sec:sec1pt3}.
In comparison with the TCDIL-collimator, the strokes of the jaws were increased to reach maximum beam impact parameters of 20.5 mm and 24.9 mm in the TCDIL-jaw and the modified version respectively. The reasons will be detailed in section ~\ref{sec:sec4}. 

\subsection{Experiment Instrumentation}
\label{sec:sec3pt3}
The experiment counted on extensive instrumentation for real-time data acquisition and monitoring of several physical values as set out in Table~\ref{tab:2}. First, it aimed at accumulating experimental input data of the beam characteristics, and especially the beam impact parameters, see section~\ref{sec:sec2pt1}. To that end, the HRM Beam TV (hereinafter ``HRM BTV'')~\cite{9,10} employs a silicon carbide (SiC) screen that is placed in the beam trajectory roughly 1m before the experiment. When the beam passes through the screen, the radiator material emits light, a camera records its intensity and software interpolates the horizontal and vertical distributions with Gaussian functions. It then gives online measurements of the beam position, as well as information about the beam spot sizes with a precision of 100 \textmu m. The Beam Pick-Up Button Position was installed to measure the relative beam bunch positions, as presented in~\cite{2}, which is related to the objective (iii) of the experiment, see section~\ref{sec:sec1pt3}.

\begin{table}[htbp]
\centering
\caption{\label{tab:2} Instrumentation overview and description of the recorded physical effects.}
\smallskip
\begin{tabular}{| l | c| l | }
\hline
\textbf{Jaw instrumentation} & \textbf{Quantity} & \textbf{Physical Effect} \\ 
\hline
Interferometer (Optic fiber based) & 3 & Bending oscillations \\ 
\hline
Linear Variable Differential Transformer (LVDT) & 1 per jaw & Quasi-static cooling \\
\hline
Platinum Resistance Thermometer (PT100) & 3 per jaw & Temperature monitoring \\
\hline
Strain gauges & 14 per jaw & Permanent strains \\
\hline
RadHard camera & 1 & Online visual observation \\
\hline
\textbf{Beam instrumentation} & & \\ 
\hline
HiRadMat Beam Television (HRM BTV) & 1 & Beam position and sigma \\
\hline
Beam Pick-Up Button Position (BPKG) & 1 & Beam bunch position \\
\hline

\end{tabular}
\end{table}

\begin{figure}[htbp]
\centering 
\includegraphics[width=1\textwidth,trim=0 0 0 0,clip]{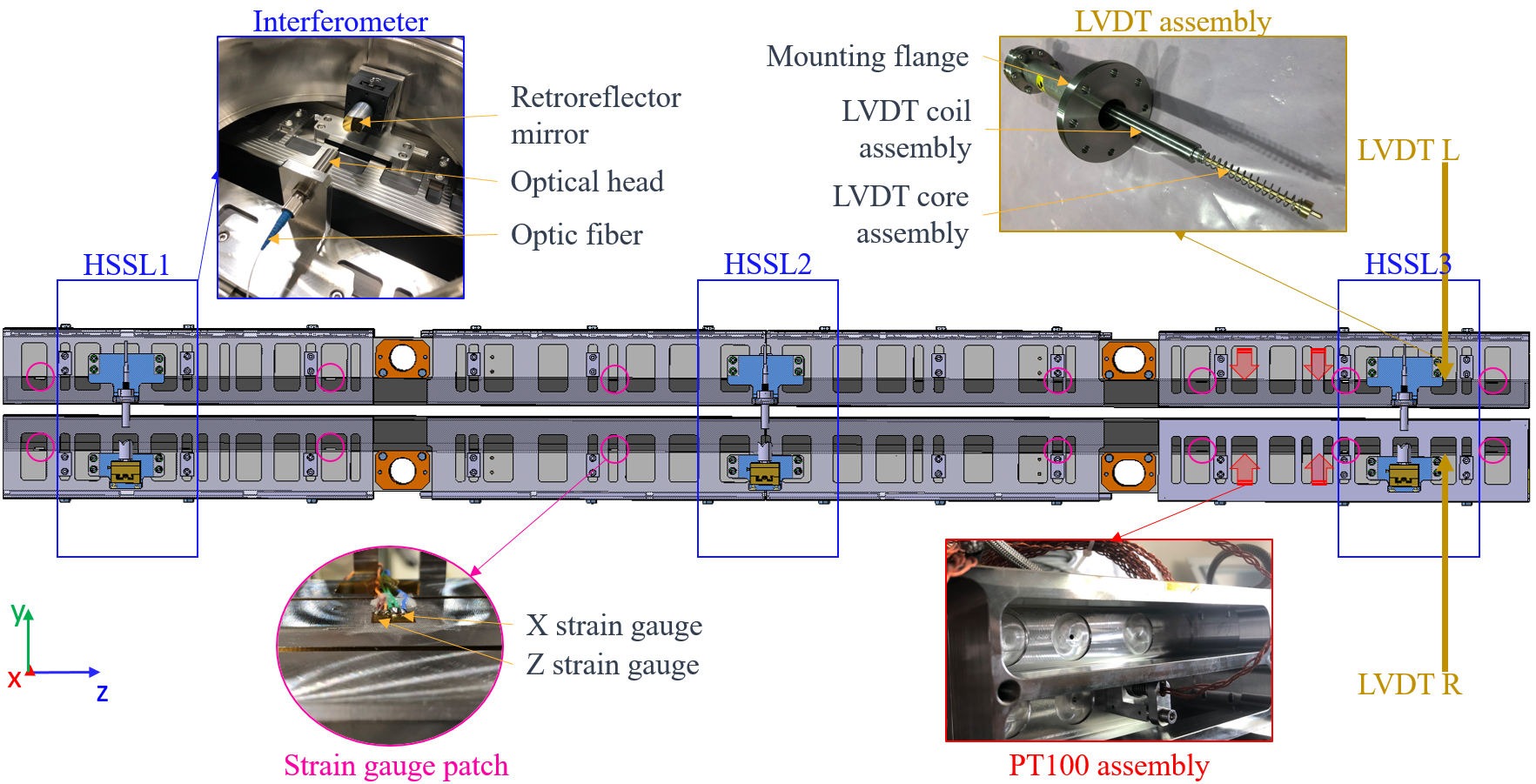}
\qquad
\caption{\label{fig:8} Detailed diagram of the jaw instrumentation used in HRMT 44. Pictures show each instrument design and the mounting position are pointed out by colored shapes.}
\end{figure}

In addition, the instrumentation aimed at accumulating experimental data to understand the jaw structural behaviour during beam impacting events and to assess online the uncertainties raised in section~\ref{sec:sec2pt3}. Figure~\ref{fig:8} presents the jaw's instrumentation by outlining its positions and giving a concise idea of the mounting. The interferometers are positioned along the two jaws in three locations, 170 mm (Identifier: ``HSSL1''), 1050 mm (Identifier: ``HSSL2''), and 1930 mm (Identifier: ``HSSL3'') from the jaw front faces, with the retroreflector mirrors mounted on the right-hand side jaw (hereinafter ``jaw (R)'') and the optical head on the left-hand side jaw (hereinafter ``jaw (L)''). HSSL stands for High Speed Serial Link, which is the communication protocol used by the interferometers. They were intended to record y-displacements, see Figure~\ref{fig:7} to offer an image of the second response phase governed by large jaw bending oscillations. A Linear Variable Differential Transformer (hereinafter ``LVDT'') (Identifier: ``LVDT\_R'' on jaw(R) and ``LVDT\_L'' on jaw (L)) is probing each jaw at 1975 mm from the front jaw face. Again, they were intended to record y-displacements, however in this case, to study the third response phase and eventually detect permanent deformation after cooling. The LVDT coil assemblies were mounted inside a vacuum flange fixed on the vacuum tank, while the LVDT core assemblies translated by maintaining contact with the backstiffener thanks to a compression spring.  Seven strain gauge patches were installed per jaw along the backstiffener face in contact with the absorbers. The patches are composed of two strain gauges measuring in the transverse and longitudinal directions, respectively in the x and z directions, to detect possible irreversible strains after cooling, which would certainly have been caused by the characteristic high compressive-to-tensile stresses of the first response phase. Finally, three Platinum Resistance Thermometer (hereinafter ``PT100'') per jaw were monitoring the jaw temperature, without being designed to measure the backstiffener's peak temperature.

The Data Acquisition System (hereinafter ``DAQ'') was configured to be able to appreciate the beam-induced jaw response by the identification of bending oscillations (with a characteristic time of a few milliseconds) and quasi-static states (with a characteristic time of a few seconds) during a total recording time of 600 s. Afterwards, an online supervisory program allowed acquisitions spread over several days to detect possible permanent deformations after cooling.

\section{Experimental procedure}
\label{sec:sec4}
HRMT 44 was focused on recreating the thermo-mechanical load of the worst case standard LIU beam accidental impact scenario by using proton beam characteristics set out in Table~\ref{tab:1}. The limited HiRadMat maximum intensity of $1.3\times 10^{11}$ protons has been compensated by increasing the beam impact parameters in order to best fit the backstiffener energy depositions. Indeed, the test bench allows to reach higher impact parameters in both jaws as explained in section~\ref{sec:sec3pt2}. 
As described in section~\ref{sec:sec2pt1}, the beam impact parameter is a sensitive beam characteristic requiring a specific experimental procedure for precise online measurements. Before the high-intensity (288 bunches) beam impacts, the jaws have been aligned to the real beam trajectory thanks to low intensity pulses (1 bunch). The procedure, so-called Beam Based Alignment~\cite{11}, suppresses mechanical errors between the HRMT BTV and the jaw active planes. In this way, the measurement error is determined by the precision of the instrument (100 \textmu m, see section~\ref{sec:sec3pt3}), the actuation system (20 \textmu m, see section~\ref{sec:sec3pt1}), and the jaw flatness (62 \textmu m for jaw (L), 103 \textmu m for jaw (R)).

\subsection{High-intensity proton beam impacts}
\label{sec:sec4pt1}
In total, four high-intensity proton beams successfully impacted the jaws with the conditions set out in Table~\ref{tab:3}. Deep (R) and deep (L) beams have been delivered after two ``grazed'' shots related to objective (iii). As requested, HRMT 44 obtained large beam impact parameters, 18.60 mm and 16.50 mm respectively, to compensate the lower maximum HiRadMat intensity in comparison to the standard LIU. The next section will evaluate the thermo-mechanical loads.

\begin{table}[htbp]
\centering
\caption{\label{tab:3} High-intensity (288 bunches) beams impacted in the HRMT-44 jaws on July, 18 2018.}
\smallskip
\begin{tabular}{| l | c| c | c | c | }
\hline
\textbf{Beam Parameters} & \textbf{Grazed 1} & \textbf{Grazed 2} & \textbf{Deep (R)} & \textbf{Deep (L)} \\ 
\hline
\textbf{Target} & jaw (R) &	jaw (R)	& jaw (R) &	jaw (L)\\
\hline
\textbf{Impact time}  &	21:03:24 &	21:39:51 &	22:44:24 &	23:38:13\\
\hline
\textbf{Intensity}	& $3.56\times 10^{13}$ & $3.57\times 10^{13}$ & $3.56\times 10^{13}$ & $3.53\times 10^{13}$\\
\hline
\textbf{Impact parameter [mm]} &	0.530 &	0.530 &	18.60 &	16.50\\
\hline
\textbf{Horizontal Spot Size\footnotemark[1]{} [mm]} & 0.450 &	0.420 &	0.420 &	0.430\\
\hline
\textbf{Vertical Spot Size\footnotemark[1]{} [mm]} & 0.380	& 0.350 & 0.350 & 0.390\\
\hline
\end{tabular}
\end{table}
\footnotetext[1]{Spot sizes measured by the HRM BTV.}

\subsection{Related energy deposition}
\label{sec:sec4pt2}
\begin{figure}[htbp]
\centering 
\includegraphics[width=0.5\textwidth,trim=0 0 0 0,clip]{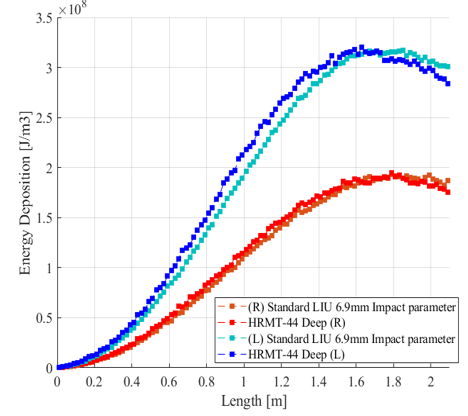}
\qquad
\caption{\label{fig:9} Energy deposition reached in the jaw backstiffener along the beam orthogonal projection in the different cases. The proton beams impacted during HRMT 44 have successfully recreated the worst LIU thermo-mechanical loads in both jaws.}
\end{figure}

FLUKA Monte Carlo simulations~\cite{i} have been employed in order to infer the energy density depositions induced by the beam impacts in the HRMT 44 jaws for the different experimental beam conditions. Figure~\ref{fig:9} shows a comparison of the energy density reached in the backstiffener along the beam orthogonal projection, and as an evidence, the worst standard LIU proton beam impacts have been successfully recreated in both jaws. The peak energy depositions would have reached $3.20\times 10^{8}$ J.m$^{-3}$ and $1.94\times 10^{8}$ J.m$^{-3}$ during deep (R) and deep (L) respectively, which approximately corresponds to maximum increases of temperature of 86 \degree C and 52 \degree C in 7.2 \textmu s.

\begin{figure}[htbp]
\centering 
\includegraphics[width=0.7\textwidth,trim=0 0 0 0,clip]{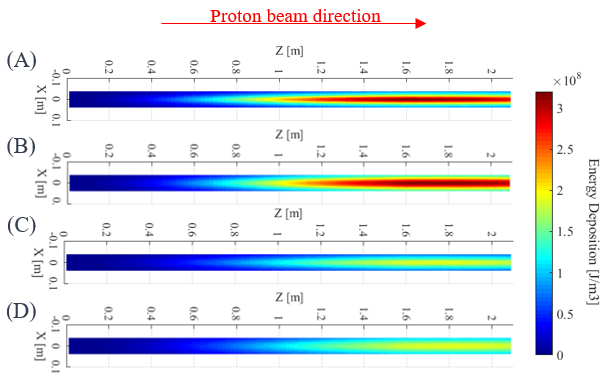}
\qquad
\caption{\label{fig:10} Energy deposition reached in the jaw backstiffener planes in contact with the absorbers. The proton beams impacted during HRMT 44 have induced more focused thermo-mechanical loads and deposited less total energy than the LIU worst case scenario. Energy deposition in Jaw (L): (A) HRMT 44 (B) LIU; in jaw (R): (C) HRMT 44 (D) LIU.}
\end{figure}

Nevertheless, a narrower heat ``footprint'' would have been induced in the HRMT 44 jaws as shown in Figure~\ref{fig:10}, together with lower total energy deposition. The coupled effect of lower beam intensities and larger impact parameters has indeed focused the thermo-mechanical loads. For example, the total energy deposition induced in the backstiffener and in the absorbers by deep (L) are 141 kJ and 301 kJ, respectively, whereas the worst standard LIU proton beam would depose 183 kJ and 489 kJ. The relative differences of 23 \% and 39 \% has limited the HRMT 44 jaw structural responses, especially by lowering the bending oscillations and quasi-static deformations. The impacted beams were however the best possible options, taking into account the current HiRadMat capabilities.

\section{Online results analysis and comparison with the numerical results}
\label{sec:sec5}
Analyzing experimental online results of the jaw structural responses is an opportunity to validate or enhance the numerical models. As described in section~\ref{sec:sec2pt2}, the preliminary analysis had highlighted two main possibilities for severely affecting the jaw flatness requirement. The following sections will present the acquired data and draw preliminary affirmations about the models' validity. It will focus on the jaw with the initial design, jaw (L), as the structural behaviors are identical to a certain scale factor related to the lower energy deposition.

\subsection{Non-induced plastic strain}
\label{sec:sec5pt1}
Fifteen strain gauges failed during the experiment, most likely because of intense vibrations and/or rapid heating that may have broken the soldered electrical connections. The others remain below 200 \textmu mm/mm with amplitudes that cannot be correlated to beam induced irreversible strains, and consequently seem to justify the elastic material assumption of the jaw structural model. This affirmation will be further investigated in the post-irradiation examinations.

\subsection{Jaw bending oscillations}
\label{sec:sec5pt2}
\begin{figure}[htbp]
\centering 
\includegraphics[width=0.75\textwidth,trim=0 0 0 0,clip]{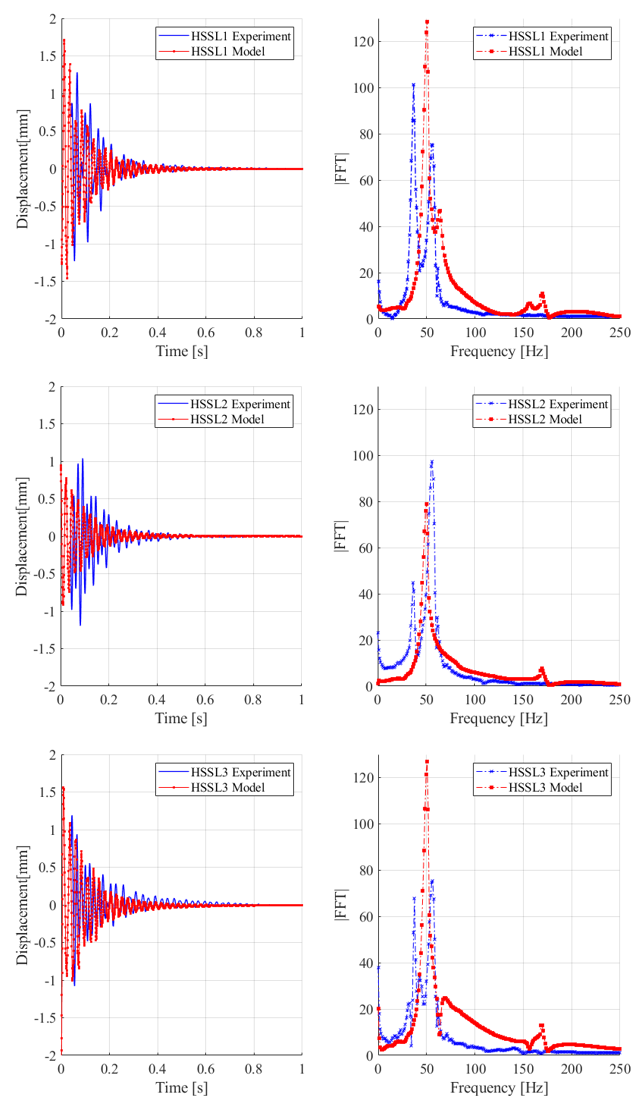}
\qquad
\caption{\label{fig:11} Time-frequency analysis of the interferometer displacement signals acquired during the high-intensity beam impact in the jaw, deep (L), and compared to the jaw structural model results. The time displacement plots, on the left hand side, show similar damping properties. The Fast Fourier Transforms give a frequency domain representation of the signals, and therefore highlight the jaw vibration modes.}
\end{figure}

Figure~\ref{fig:11} presents a time-frequency analysis of the interferometer displacement signals acquired during the high-intensity beam impact deep (L). Signal losses occurred in the very first instants up to 40 ms, certainly due to some internal limitations of the instruments, and have disrupted the absolute displacement data. Thus, some post-processing has been performed by cleaning the signal losses and offsetting the mean displacement to zero in order to clear the incorrect constant term. 

The damping properties of the jaw structural response can be appreciated, and have been used to update the first estimate of the friction coefficient at the interfaces between the guiding plates and the cylindrical shafts. The iterative approach has converged to \textmu\ = 0.1 to finally give good agreement between the experimental and numerical results. It also confirms that the jaw bending oscillations are completely damped 1 s after the proton beam impact.

On the other hand, the Fast Fourier Transform offers a mathematical tool to identify the dominant vibration modes in the frequency domain. The experimental response is dominated by two frequencies at 36 Hz and 56 Hz, observable through the three installed interferometers. The 56 Hz-frequency mode is correctly predicted by the numerical model, and as discussed in section~\ref{sec:sec2pt2}, corresponds to the jaw fundamental. However, the 36 Hz-frequency mode remains a mystery at this time. It cannot be accounted by a jaw vibration mode, so the study may further investigate solid modes related to the jaw support and the actuation system.

\subsection{Residual deformations after cooling}
\label{sec:sec5pt3}
\begin{figure}[htbp]
\centering 
\includegraphics[width=0.7\textwidth,trim=0 0 0 0,clip]{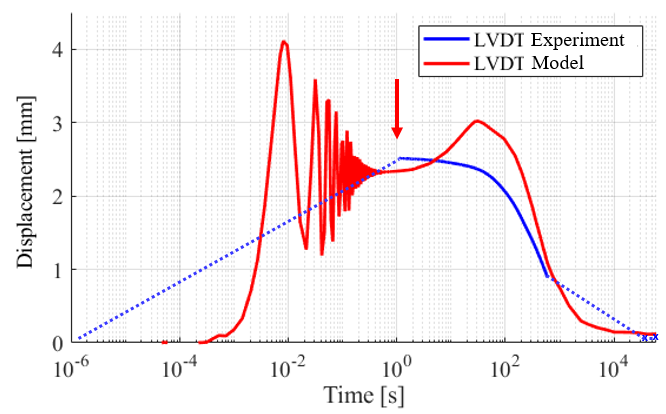}
\qquad
\caption{\label{fig:12} The LVDT\_L signal acquired during the high-intensity beam impact, deep (L), is plotted in solid blue, whereas the spread acquisitions during cooling are marked by a blue cross. The y-displacement calculated by the numerical model at the probing spot is plotted in solid red. Some inconsistencies are visible in the first instant of the third response phase.}
\end{figure}

Figure~\ref{fig:12} shows the signal recorded by LVDT\_L during the high-intensity beam impact deep (L) during 600 s, which corresponds to the DAQ recording time. Some post-processing has been required to clean the signal before 1 s as large disruptions with non-physical amplitudes occurred. Because of the jaw vibration, the LDVT core was certainly interfering with the LVDT coil assembly. The plot has been completed by two spread acquisition points at 40,000 s and 53,000 s. In addition, the y-displacement calculated by the numerical model at the probing spot is plotted in red. The characteristic jaw bending oscillations of the second response phase are clearly visible, as explained in section~\ref{sec:sec5pt2}. 

After that phase, as discussed in section~\ref{sec:sec2pt2}, the jaw acquires a convex quasi-static geometry caused by thermal strains concentrated in the absorbers and in the lower volumes of the backstiffener. One second after the proton beam impact, the signals are comparable in terms of amplitude, 2.515 mm and 2.341 mm for the experimental and numerical results respectively. Afterwards, the curves diverge as the model predicts a displacement increase until 30 s and then a continuous decrease, whereas a continuous displacement decrease has been observed during the experiment. Various phenomena could explain such an inconsistency, especially the wrong prediction of the sticking-slipping states at the boundary interfaces between the guiding plates and the cylindrical shafts, but it requires further investigation. Approximately 4 days were necessary for the jaw to cool down at the initial temperature conditions. At the end, permanent displacements have been detected in both signals. The amplitudes are comparable, 58 \textmu m and 118 \textmu m according to the experimental and numerical results respectively.

\section{Post-irradiation examinations}
\label{sec:sec6}
After the experiment was completed, the HRMT 44 test bench was stored for two months in a dedicated area to reach acceptable residual radioactive dose rates. The following post-irradiation examinations first aimed at comparing the jaws' geometry before and after the high-intensity proton beam impacts by means of metrology measurements. It aspires to consolidate the online results analysis in section~\ref{sec:sec5}, more specifically to confirm both the absence of plastic strains in the backstiffener and the slight permanent jaw deformations after cooling. On the other hand, a visual observation was required to confirm that no damage had been induced in the absorber blocks in Novoltex\textregistered\ Sepcarb\textregistered\ 054-62.

\subsection{Metrology measurements}
\label{sec:sec6pt1}
\begin{figure}[htbp]
\centering 
\includegraphics[width=1\textwidth,trim=0 0 0 0,clip]{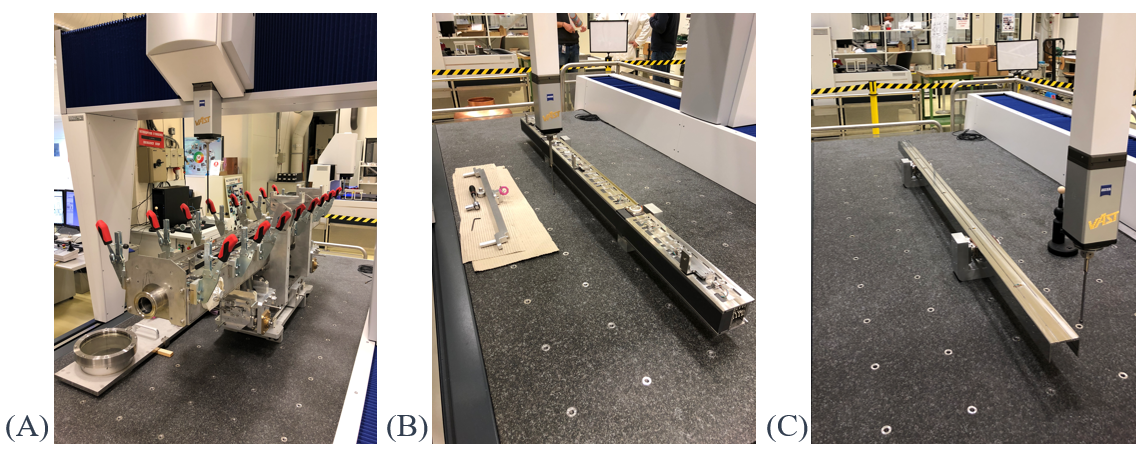}
\qquad
\caption{\label{fig:13} The different configurations used for metrology measurements. (A): Jaws in the HRMT 44 test bench. (B): Jaws hold up on two supports similar to free mounts. (C): Bare backstiffener hold up on two supports similar to free mounts.}
\end{figure}

As shown in figure~\ref{fig:13}, metrology measurements have been performed in different configurations, starting from the experimental assembly with the jaws still mounted inside the HRMT 44 test bench to a freely supported jaws, and finally to a freely supported bare backstiffener. A Zeiss Prismo Ultra MMT machine with a range of 2400 mm $\times$ 1200 mm $\times$ 1000 mm and a precision of 1.2 + L/500 \textmu m was used for the measurements at an ambient temperature of 20 $\pm$ 1 \degree C. The analysis also compares the two jaw designs to examine the benefits brought by the upgraded version.

\subsubsection{Jaws in the HRMT 44 test bench}
\begin{figure}[htbp]
\centering 
\includegraphics[width=1\textwidth,trim=0 0 0 0,clip]{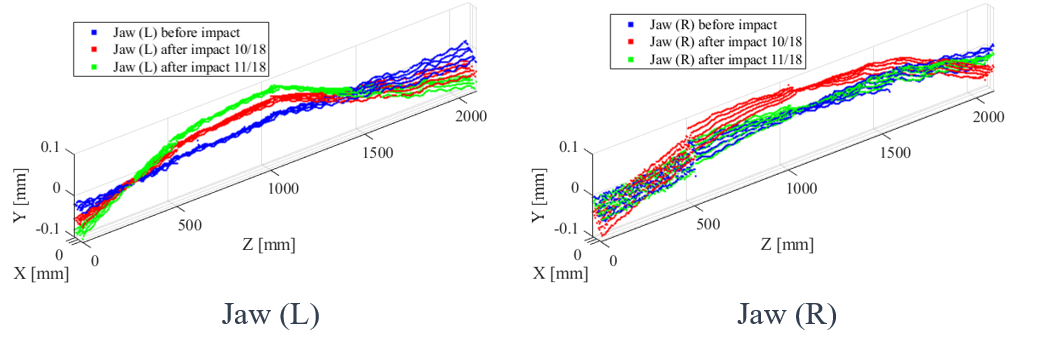}
\qquad
\caption{\label{fig:14} Visual representation of the jaw flatness characteristics before and after the high-intensity proton beams, while the jaws were installed in the HRMT 44 test bench. Two post-measurements have been performed within a period of 1 month, and show how the jaw geometry is sensitive to proton beam impacts, as well as transportation.}
\end{figure}

Figure~\ref{fig:14} presents the metrology measurements carried out while the jaws were still installed in the HRMT 44 test bench and gives a visual representation of the jaw forms thanks to flatness geometric characteristics. After the high-intensity proton beam impacts, two measurement runs were performed within a period of 1 month, while a CERN internal transport occured in-between. Both the TCDIL jaw, jaw (L) on the left-hand side, and the upgraded version, jaw (R) on the right-hand side, had acquired a more convex geometry, which corresponds to the LVDT online results. The flatness values were respectively of 62 \textmu m and 103 \textmu m before closing the vacuum vessel, and had reached 124 \textmu m and 168 \textmu m just after opening it in October 2018. Surprisingly, noticeable changes had been detected in November, as jaw (L) acquired an even more convex geometry with a flatness value of 184 \textmu m, and jaw (R) straightened its form up by reaching a flatness value of 97 \textmu m (very similar to the value measured before the high-intensity proton beam impacts). 

These measurements have highlighted how the jaws can be affected by shocks and vibrations generated by transportation as well. It seems that the geometry is sensitive to the connection between the cylindrical shafts and the jaw guiding plates, which by a stick-slip phenomenon can lead to permanent deformations. 

\subsubsection{Freely supported jaws}
\begin{figure}[htbp]
\centering 
\includegraphics[width=1\textwidth,trim=0 0 0 0,clip]{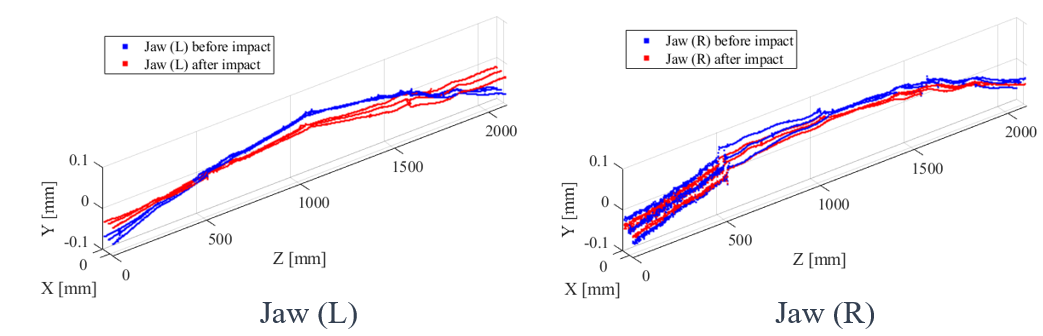}
\qquad
\caption{\label{fig:15} Visual representation of the jaw flatness characteristics before and after the high-intensity proton beams, while the jaws were freely supported. The proton beam impacts have straightened the jaw forms up.}
\end{figure}

As a second step, the jaws were removed from the HRMT 44 test bench, and measured with ``sliding'' contact planes on the guiding plates. The jaw forms are presented in Figure~\ref{fig:15}. Both jaws became straighter as the flatness values go from 134 \textmu m and 133 \textmu m, respectively for jaw (L) and jaw (R) before the high-intensity proton beam impacts, to 62 \textmu m and 106 \textmu m afterwards.

\subsubsection{Freely supported backstiffener}
\label{sec:sec6pt1pt3}
\begin{figure}[htbp]
\centering 
\includegraphics[width=1\textwidth,trim=0 0 0 0,clip]{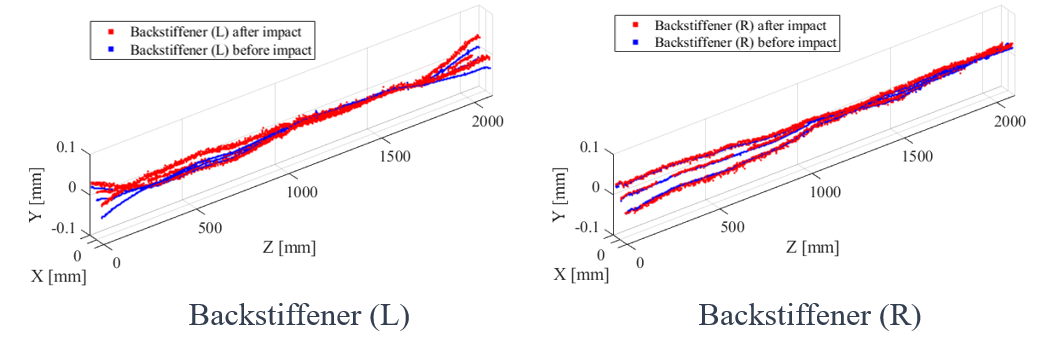}
\qquad
\caption{\label{fig:16} Visual representation of the backstiffener flatness characteristics before and after the high-intensity proton beams. The proton beam impacts have not altered the forms, which seems to validate the absence of irreversible strains and therefore, the elastic material assumption of the numerical model.}
\end{figure}

As a final step, both jaws were disassembled and metrology measurements have been performed on the backstiffeners, again with a free mount. The flatness visual representations in Figure~\ref{fig:16} clearly confirm that the backstiffeners were not affected by the proton beam impacts as the flatness values are very stable: from 79 \textmu m and 66 \textmu m for backstiffener (L) and backstiffener (R) before, to 60 \textmu m and 56 \textmu m after, respectively. These measurements seem to validate the model's elastic assumption.

\subsection{Absorber visual observations}
\label{sec:sec6pt2}
\begin{figure}[htbp]
\centering 
\includegraphics[width=1\textwidth,trim=0 0 0 0,clip]{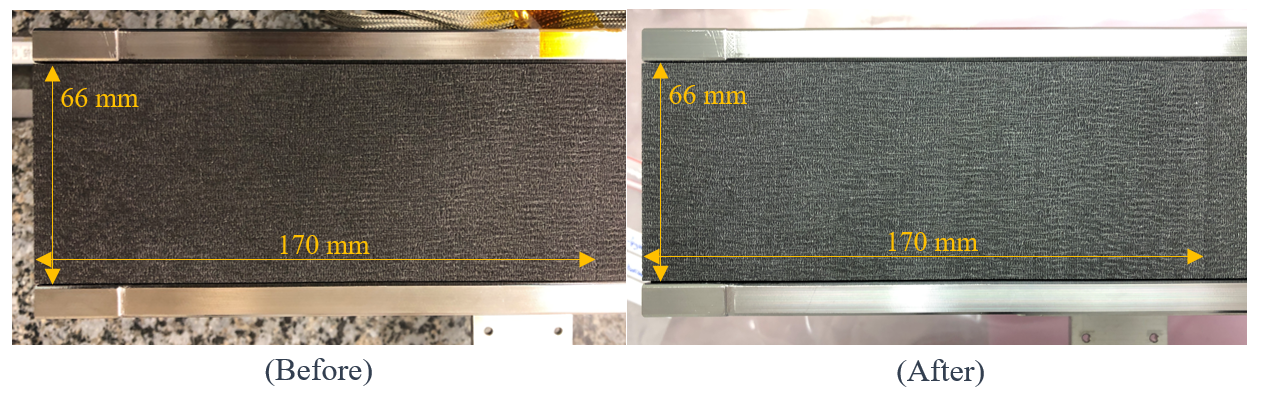}
\qquad
\caption{\label{fig:17} Picture of the upstream Novoltex\textregistered\ Sepcarb\textregistered\ 054-62 absorber before (May 2018) and after (January 2019) the HRMT 44 high-intensity proton beam impacts. No defect is visible.}
\end{figure}

A visual observation of the three Novoltex\textregistered\ Sepcarb\textregistered\ 054-62 absorbers was performed before and after the HRMT 44 high-intensity proton beam impacts. Figure~\ref{fig:17} shows pictures of the upstream absorber, in which the most critical thermo-mechanical loads occurred. The comparison has not brought any defect, cracks or visible deformations to light, and seems to confirm the material integrity after the beam Grazed 1 and Grazed 2 impacts. The study is detailed in~\cite{5}, in which the induced loads are evaluated through a numerical model using the experimental beam data, especially the beam spot size and the beam impact parameter, and in which the results are compared to future possible LIU scenarios.

\section{Conclusion}
\subsection{Findings}
The HRMT 44 test bench has successfully been impacted by high-intensity proton beams at CERN's HiRadMat facility to recreate in two jaws the thermo-mechanical loads of the worst accidental case scenario considered by the LIU project. As introduced, the experiment's objectives were to validate the TCDIL collimator jaw design, challenging in view of the 0.200 mm flatness tolerance requirement and to enhance the jaw numerical model.

Firstly, the absence of relevant plastic strains, and consequently the model's elastic assumption, has been confirmed by both the experimental online results and the post-irradiation examinations. They have proved that the jaw core part, the backstiffener, is not altered by the worst proton beam impacts, and this in spite of the large dynamic stresses induced by the thermal burst. 

Then, the study has highlighted how the jaw geometry, while installed in the test bench and by extension, in a TCDIL collimator, is sensitive to the connection between the cylindrical shafts and the jaw guiding plates, regardless of the jaw design. At that time, frictional contacts model the connection with a friction coefficient based on the analysis of the jaw response damping properties. Numerical predictions are however extremely challenging as they rely on a precise description of stick-slip phenomena, which actually depends on many external parameters. Experimental estimations to further validate this hypothesis are consequently suggested.

On the one hand, it seems that even less critical beam impacting events would induce convex permanent deformations, with an alteration of the flatness value up to 70 \textmu m. On the other hand, handling activities for transportation, installation or maintenance, have the capacity, as well, to alter randomly the jaw characteristics. Depending of the shock and/or vibration direction, the flatness can either be improved or degraded to an extent of $\pm$ 70 \textmu m.

The manufacturing of the TCDIL collimators will be soon completed, and installation in the SPS-to-LHC transfer lines is foreseen for 2020. 

\subsection{Further studies}
Some open points have nevertheless been highlighted during this study. For example, an unexpected vibration mode at a frequency of 36 Hz has been detected through the interferometer online results, whereas the jaw fundamental frequency is 56 Hz. The quasi-static jaw response during cooling seems, in addition, to require a deeper understanding of the stick-slip phenomena at the connection. 

Finally, a complex phenomenon has been highlighted by the metrology measurements of the jaw when removed from its supports. The geometries were indeed straightened, even if the backstiffener integrity were not affected.

\newpage
\acknowledgments
The authors want to acknowledge all the personnel from the different CERN services involved in the HRMT 44 experiment for the support in the execution of design (EN-MME and EN-STI), manufacturing (EN-MME and EN-STI), assembly (EN-STI), installation (EN-HE, EN-EA and EN-STI), execution (BE-OP) and post-irradiation examination activities (HSE-RP-AS and EN-MME-MM). Their collaborative efforts have made the experiment a success. In addition, the authors would like to particularly thanks F. Harden (EN-EA) for her essential support in the coordination of the HiRadMat facility and C. Bracco (TE-ABT) for her effort carrying out the Beam Based Alignment of the jaws. Finally, the authors thank warmly Louisa Catherall for the English proofreading of this paper.

\end{document}